\magnification=1200
\nopagenumbers
\centerline{\bf{Quantum fluctuations of some gravitational waves $^{*}$ }}
\vskip .7truein
\centerline{ Y.Enginer $^{*}$, M.Horta\c csu$^{*+}$, R.Kaya$^{+}$,
N.\" Ozdemir$^{*+}$
, K. \" Ulker$^{*}$, B.Yap\i \c skan$^{*+}$}
\vskip .6truein
\centerline {$^{*}$ Physics Department, Faculty of Sciences and Letters,
ITU, 80626, Maslak, Istanbul, Turkey}
\vskip .2truein
\centerline {and}
\vskip .2truein
\centerline{$^{+}$ T\" UBiTAK Feza G\" ursey Institute, Kandilli, Istanbul,
Turkey}
\vskip 1.3truein
\centerline {Abstract}
We review our previous work on the calculation of
the vacuum expectation value of the stress-energy tensor for
a scalar particle in the background metric of
different types of spherical impulsive, spherical shock and plane impulsive
gravitational waves.  We get non-zero vacuum fluctuations only if we
use the spherical impulsive wave  as the background.  We take a detour in
de Sitter space for regularizing our result both in the ultraviolet and
the infrared regions.

\def\kutu{{\rlap{$\sqcup$}\sqcap}}
\vskip 1.7truein
\noindent 
* This work is dedicated to the 225'th anniversary of I.T.U.
\baselineskip=18pt
\footline={\centerline{\folio}}
\pageno=1
\vfill\eject
\noindent
{\bf{INTRODUCTION}}

\noindent
We know that the methods for the quantization of Einstein's theory of
gravititation are not at their final stage at the moment.  There are many
models which incorporate this theory into the models for elementary
particles.  One can cite all the string, membrane, M models, etc.
which are extensively studied $^{/1}$.  The final word is not still said in
these models.

In the absence of a generally accepted method for the quantization of
gravitation, semi classical methods were used to extract information
about the theory, which were very useful in the absence of a full
quantization.  We can calculate the fluctuations in the stress-energy tensor
of a particle that propagates through universes described by different
metrics.  Extensive work in this direction
was done in the seventies stressing the phenomena of particle production
in these metrics    . This work is  described in the
books written by Birrell and Davies $^{/2 }$, Fulling $^{/3}$ and Wald
$^{/4}$ .  Essentially we are confronted
by a problem of a particle in an external field, as the metric contribution
can be written  as an external potential.  The n-point functions where
n exceeds two can be found by a simple application of the Wick's theorem
in quantum field theory.  Kuo and Ford show the calculation of the four-point
function, thereby variance in the stress-energy tensor using this
idea $^{/5}$.

In the eighties and nineties people applied the same method to problems
in the presence of gravitational waves, or to the effects of different
topological structures in the space-time manifold.  Vacuum fluctuations
in the presence of cosmic strings is an example of such work $^{/6  }$.

We were interested in the same problem in the presence of spherical
waves created by the snapping of cosmic strings .  We calculated the
vacuum-expectation-value, VEV, of the stress-energy tensor, $<T_{\mu \nu}>$,
for several types of spherical impulsive and shock waves,  described
by different warp functions.  There is a
general argument $^{/7 }$ backed up by a explicit calculation $^{/8   }$
which prohibits the polarization of the vacuum in the presence of plane
waves.  It was not known whether the same obstruction prevailed in the
spherical wave case, though; so, the expression for $<T_{\mu \nu}>$
was calculated $^{/9}$ for special cases of the Nutku-Penrose metric $^{/10}$.
We found that although a finite expression can not be extracted
in the first order calculation, the second order calculation indeed gives
a finite result .  Here we had to take a detour in the de Sitter space and
take the appropriate limit to go back to the Minkowski space at the end
of the calculation.

The same method was applied $^{/11}$ to the spherical shock
wave given by Nutku $^{/12}$.  In this case we were not able to extract
a finite expression even in second order.

We, then, applied our method to the impulsive plane waves.  Relying on the
Deser-Gibbons work $^{/7,8  }$, we were not anticipating a finite result
at this point.  It seems that the method we proposed for this calculation
passed this test.  We could not get a finite contribution to the vacuum
fluctuations if we had a plane impulsive wave propagating in the Minkowski
space.  If we have an impulsive plane wave in the de Sitter universe, though,
we get a finite result which is proportional to the cosmological constant.
We also find that the presence of impulsive pp waves does not change
this result.

In Section II we first calculate $<T_{\mu \nu}>$ for an impulsive wave
in first order perturbation theory for a special form of the warp function.
In Section III we do the similar calculation for a specially simple form
of the same function in second order perturbation theory.  Section IV is
devoted to the  similar problem in the de Sitter background.  Here we show how
one can extract a finite expression for $<T_{\mu \nu}>$ first by performing
the calculation in de Sitter space and then taking the appropriate limit to
land in Minkowski space.  In Section V we do the similar calculation for the
shock wave and point to the absence of a finite term in the expression
for $<T_{\mu \nu}>$ even in second order  in perturbation theory.
In Section VI the similar calculation is repeated for an impulsive plane
wave and it is shown how the proposed method fails to extract a finite expression
for the vacuum expectation value  of the stress-energy tensor in the
background metric of a plane wave.  This result is in agreement with
the result given by Deser and Gibbons $^{/7,8}$.  We also show
how this calculation can be extended to pp waves and give the result
for this case.  We end up with few concluding  remarks.
\bigskip
\noindent
{\bf{II. Spherical impulsive wave: first-order calculation}} 

\noindent
The Nutku-Penrose metric $^{/10 }$ for spherical gravitational waves formed
as a result of snapped cosmic strings is given by
$$ ds^2= 2 dudv-u^2|d\zeta+v\Theta(v) f(\overline{\zeta})
d \overline{\zeta} |^2 .   \eqno {1} $$
Here $v$ is the retarded time, $u$ is a Bondi-like luminosity distance,
$\zeta$ is the angle of stereographic projection on the sphere.
$f$ is the Schwarzian derivative of an arbitrary holomorphic function
$h(\zeta)$ which describes the way the cosmic string is snapped and is called
the warp function.  Only one component of the Weyl tensor is not zero.
It is given by
$$ \Psi_{4} = {{\delta(v)} \over {u}} f, \eqno {2} $$
exhibiting the characteristic behaviour of an impulsive wave.
We expect similar behaviour for the vacuum expectation value of the
stress-energy tensor ,$<T_{\mu \nu} >$, for a massless scalar particle
propagating in the background metric of such a wave if we get any finite
value for this expression.

We choose a particular form for the warp function $h$ to perform an explicit
calculation.  Our experience with the different forms we chose $^{/9 }$
shows that the form  of the expression for $<T_{\mu \nu}>$ is not sensitive
to the actual choice for $h$.  The presence of the Dirac delta function
and the presence of a finite expression depends only on the general
properties of the metric.  The functional form of the warp function only
effects the dependence on $\zeta,\overline{\zeta}$.  This fact allows
us to choose  a special form for this function that will make our
calculations as simple as possible.

Our first choice corresponds to a "rotating" string and is given by
$$ h=(\zeta)^{1+i\epsilon} \eqno {3} $$
which gives
$$f(\zeta)= {{-i\epsilon} \over {\zeta^2}} \eqno {4} $$
in first order in $\epsilon$.  With this choice we rotate the shear of the
metric by $\pi/2$ compared with the real exponent case.  We take the
rotating string  because the calculations are somewhat simpler in this
case.  Here $\epsilon$ is a small number which can be positive or negative,
and will be used as the perturbative expansion parameter.

First we go to real coordinates
$$\zeta= {{1}\over {\sqrt {2} }} (x+iy) \eqno {5} $$
and write the metric, up to first order in $\epsilon$, for $v>0$,
$$ds^2=2dudv-\left( u^2-{{8\epsilon vuxy} \over {(x^2+y^2)^2}}\right) dx^2
-\left(u^2+{{8\epsilon vuxy}\over {(x^2+y^2)^2}} \right) dy^2
-{{8\epsilon vu} \over {(x^2+y^2)^2}} (x^2-y^2) dxdy. \eqno {6} $$
The d'Alembertian operator $\kutu$ times $\sqrt{-g}$
, written in the background of this metric reads
$$\sqrt{-g} \kutu=$$
$$2u^2 {{\partial ^2}\over {\partial u \partial v}}+
2u {{\partial} \over {\partial v}} -{{\partial ^2 } \over {\partial x^2}}
-{{\partial ^2} \over {\partial y^2}}-{{8\epsilon v} \over {u(x^2+y^2)^2}}
\left [ xy \left( {{\partial ^2} \over {\partial x^2}} -{{\partial ^2}
\over {\partial y^2}} \right) -(x^2-y^2) {{\partial ^2} \over {\partial x
\partial y}} \right]. \eqno {7} $$

For technical reasons we first couple a massive scalar field to this metric
and take the limit mass going to zero at the end of the calculation.
We use conformal coupling. The Ricci scalar is zero in the Minkowski
case.  The form of this coupling only matters in the de Sitter case.

We first compute the Green's function by summing the eigenvalues of the
operator
$$ G_F = \sum_{\lambda} {{\phi_{\lambda} (x) \phi_{\lambda}^{*} (x')}
\over {\lambda}} \eqno {8} $$
where
$$ L\phi_{\lambda}=\lambda \phi_{\lambda} ,\eqno {9} $$
$$L = \sqrt{-g} [\kutu+m^2]. \eqno {10} $$
We calculate $<T_{\mu \nu}>$ by taking the appropriate derivatives of $G_F$.
Since our computation is in first order in $\epsilon$, we expand
$L,\phi_{\lambda}$ and $\lambda$ and write
$$(L_0+\epsilon L_1)(\phi_0+\epsilon \phi_1+...)=
(\lambda_0+\epsilon \lambda_1+...)
(\phi_0+\epsilon \phi_1+...) . \eqno {11} $$
This gives us
$$ L_0 \phi_0 = \lambda_0 \phi_0 , \eqno {12} $$
$$L_1 \phi_0+L_0 \phi_1= \lambda_1 \phi_0 + \lambda_0 \phi_1,
\eqno {13} $$
where
$$L_0=\left(2u^2{{\partial ^2} \over {\partial u \partial v}}+2u{
{\partial} \over {\partial v}}-{{\partial ^2} \over {\partial x^2}}
-{{\partial ^2} \over {\partial y^2}} +2 m^2 u^2 \right), \eqno {14} $$
$$ \phi_0= {{ e^{iRv} e^{ik_1x} e^{ik_2y} e^{{{-iK}\over {2uR}}}
 e^{{{im^2u}\over {R} }}}\over {u \sqrt{|2R|} (2 \pi)^2}} \eqno {15} $$

$$\lambda_0=k_1^2+k^2_2-K \eqno {16} $$

$$L_1={{8v}\over {u}} \left[{{xy}\over {(x^2+y^2)^2}}
\left({{\partial ^2} \over {\partial x^2}}-{{\partial^2} \over
{\partial y^2}}\right)
- {{x^2-y^2}\over {(x^2+y^2)^2}}{{\partial ^2} \over {\partial x \partial y}}
\right], \eqno {17} $$

$$\lambda_1= (\phi_0,L_1 \phi_0) =0 . \eqno {18} $$

\noindent We write
$$ \phi_1= \phi_0 f. \eqno {19} $$
The form of the operator $L_1$ given in eq. 17 suggests the ansatz
$$ f=vf_1(x,y,u)+f_2(x,y,u) \eqno {20} $$
for $f$.  This ansatz gives us two coupled equations
$$L' f_1= {{8}\over {u}} \left[ {{xy}\over{(x^2+y^2)^2}}(-k_1^2+k_2^2)
+{{(x^2-y^2)k_1k_2} \over {(x^2+y^2)^2}} \right] ,\eqno {21} $$
$$L' f_2= -2u^2 {{\partial f_1}\over { \partial u}} -\left({{iK} \over {R}}
+{{2im^2u^2} \over {R}} \right) f_1, \eqno {22} $$
where
$$L'=2u^2iR {{\partial} \over {\partial u}}-2i\left(k_1{{\partial}\over
{\partial x}}
+k_2{{\partial} \over {\partial y}} \right) -{{\partial ^2} \over
{\partial x^2}}
-{{\partial ^2} \over {\partial y^2}} . \eqno {23} $$
Here we take mass only as an infrared regulator and from this point on
equate it to zero everywhere
except in the zeroth order solution $\phi _{0}$.  We will take the limit
where it goes to zero at the end of our calculation.

At this point we see that it is convenient to make the change of variables and
use
$ z=x+iy, \overline{z} =x-iy,$ and $ s={{1}\over {u}}$.  In terms of these
variables we get
$$L'=-2iR{{\partial}\over {\partial s}} -2i\left[ (k_1+ik_2){{\partial}
\over {\partial z}}
+(k_1-ik_2){{\partial} \over {\partial {\overline {z}}}} \right]-
 4{{\partial^2}\over{\partial z \partial {\overline{z}}}}\eqno {24} $$
$$ L'f_1=- 2is\left( {{(k_1+ik_2)^2}\over {z^2}}-{{(k_1-ik_2)^2} \over
{\overline{z}^2}}
\right) \eqno {25} $$
with the immediate solution
$$f_1=R(ln{z}-ln{\overline{z}})-\left({{ (k_1+ik_2)s}\over {z}}
-{{(k_1-ik_2)s} \over{\overline{z}}} \right) \eqno  {26} $$
$$f_2=-{{Ks}\over{2R}} ({{1}\over {z}}-{{1}\over{\overline{z}}})-{{i}\over{2}}
(ln z-ln{\overline{z}})
+K\left({{zlnz-z} \over{k_1+ik_2}} -{{\overline{z} ln{\overline{z}}
-\overline{z}}\over{ k_1-ik_2}}\right) .\eqno {27} $$
If we go back to the real variables the solutions read
$$f_1=2Ri tan^{-1} {{y}\over{x}}-2is {{k_2x-k_1y}\over{x^2+y^2}} \eqno {28} $$
$$f_2=-({{iKs}\over{R}}-{{1}\over{2}}) tan^{-1} {{y}\over{x}}
+{{iK}\over{k_1^2+k_2^2}}\left( 2(k_1 x+k_2y) (tan^{-1}{{y}\over{x}} -y)
\right.$$
$$\left. +(k_1y-k_2x)\left((ln x^2+y^2) -2x\right)\right) \eqno {29} $$

To form the Green's function $G_F$, we have to calculate
$$G^{(1)}_F= \int_{- \infty}^{\infty} dk_1 \int_{-\infty}^{\infty} dk_2
\int_{-\infty}^{\infty} dK \int_{-\infty}^{\infty} dR
{{\phi_0^{\lambda}(x) \phi_1^{ \lambda *} (x')+ \phi_0^{\lambda *}(x')
\phi_1^{\lambda}(x) }\over { K-k_1^2-k_2^2}} ,\eqno {30} $$
where $\phi^{\lambda}_0 $ and $ \phi ^{\lambda}_1 $ are the zeroth and
the first order solutions respectively.
We have to multiply $f_1$ and $f_2$ by the operator
$$ {\bf{L}}= i\int_{0}^{\infty} d\alpha
\int_{-\infty}^{\infty} dk_1 \int_{-\infty}^{\infty} dk_2
\int_{-\infty}^{\infty} dK  {{1}\over {(8\pi^2)(2\pi)^2uu'}}
\int_{-\infty}^{\infty} {{dR}\over{2|R|}} e^{{-iK(s-s')}\over {2R}} $$
$$ \times e^{iR(v-v')} e^{ik_1(x-x')} e^{ik_2(y-y')}
e^{i\alpha(K-k_1^2-k_2^2)} e^{{{im^2} \over {R}} ({{1}\over{s}}-{{1}
\over{s'}})} \eqno {31} $$
At this point we are looking for finite contributions to the $<T_{\mu \nu}>$.
To calculate them we have to first calculate the Green's function, and
then differentiate it to obtain the vacuum expectation value of the
stress- energy tensor.  Since we expect a non zero contribution, if any,
only from  $<T_{vv}>$, taking the coincidence limit in $x$ and $y$
before we calculate $G_F$ will not change the end result because we have
to differentiate with respect to $v$ and $v'$ only.  If these terms are zero
in the coincidence  limit in $x$ and $y$ for the Green function, they will
be also zero for $<T_{\mu \nu}>$.

A straightforward calculation shows that if we take the coincidence limit
for $x$ and $y$, i.e. take $x=x',y=y'$, we find
$$G_F^{(1)}= [A{{(\Theta(v)+\Theta(v'))}\over {(u-u')(v-v')}}
+B{{(v\Theta(v)-v' \Theta(v'))}\over {(u-u')(v-v')^2}}]
tan^{-1}{{y}\over{x}} .\eqno {32}$$
Here $A,B$ are numerical constants. We could not extract a finite
portion out of these expressions
in the coincidence limit.

One can show that if we go one order higher,
we may get finite contributions if we take a detour in de Sitter space.
Since this calculation is quite involved, we first use a simpler
form of the warp function and  demonstrate how this method works.
\bigskip
\noindent
{\bf{III. Spherical impulsive wave: second-order calculation}}

\noindent
In this section we take the warp function $h= e^{\alpha \zeta} $, which
results in $f=-{{\alpha^2}\over {2}}$, resulting in a metric
$$ds^2=2 dudv-{{1}\over{4}}\left(dx^2(2u-v\alpha^2 \Theta (v))^2
+dy^2(2+v\alpha^2 \Theta(v))^2\right). \eqno {33} $$
The d'Alembertian operator in this metric reads
$$\kutu={{1}\over{\sqrt{-g}}} \partial_{u}(g^{\mu \nu} \sqrt{-g} ) \partial_v
=
2\partial_{\mu}\partial_{\nu} + \left( {{1}\over {u-{{\alpha^2 v}\over {2}}}}
+{{1} \over {u+{{\alpha^2 v} \over {2}}}} \right) \partial_v $$
$$+ {{\alpha^2}\over{2}} \left( {{1}\over{u+{{\alpha^2 v}\over {2}}}}
-{{1} \over {u-{{\alpha^2 v} \over {2}}}} \right) \partial_u
-{{1}\over {(u-{{\alpha^2 v}\over {2}})^2}} \partial_x^2
-{{1}\over {(u+{{\alpha^2 v} \over {2}})^2}}
\partial_y^2. \eqno {34} $$
for the exact operator.
We multiply this expression , eq. {34}, by $\sqrt{-g}$ which is
equal to $u^2$ and expand the operator up
to second order in $\alpha^2$:
$$L^{II}=2u^2 \partial_{u} \partial_{v} +2u \partial_{v} -\partial_{x}^2
-\partial_{y}^2-{{\alpha^2v}\over{u}}(\partial_x^2-\partial_y^2)-\alpha^4
\left( {{v}\over{2}} \partial_u -{{v^2}\over{u}} \partial_v+{{3}\over {2}}
{{v^2}\over{u^2}}(\partial_x^2+\partial_y^2) \right). \eqno {35} $$

Here, again, we will first compute the Green's function for this operator,
and then differentiate it to get the vacuum expectation value of the
stress-energy tensor.  We obtain the same zeroth order solution as given
in eq. 15 and make the same ansatz
$\phi^{(1)}=\phi^{(0)} f$ since the first order operator is of the same form
as previous case.
We see that $f$ just modulates the zeroth-order solution, and does not
essentially change it in a radical manner.

$f$ obeys the differential equation
$$L_2 f ={{ v}\over {u}} (k_2^2-k_1^2) \eqno {36} $$
where $L_2$ is defined as
$$L_2=\left( -2iR {{\partial} \over { \partial s}}
-2i\left( k_1{{\partial} \over {\partial x}}
+k_2{{\partial}\over {\partial y}} \right)
-{{\partial^2}\over {\partial x^2}}
-{{\partial^2}\over {\partial y^2}}-
2{{\partial ^2} \over {\partial s \partial v}}
+{{iK}\over {R}}{{\partial} \over {\partial v}} \right) \eqno {37} $$
Here the modes $k_1,K_2,R,K$ are as given in eq. 16, $s={{1}\over{u}}$.
We make the ansatz for $f$  similar  to the one given in eq. 20.
$f=vf_1(s,x,y)+f_2(s,x,y)$.
This yields equations similar to the ones given in eq.s 21 and 22.
We get
$$L'f_1=s(k_2^2-k_1^2)  ,\eqno {38} $$
$$L'f_2=\left(2 {{\partial} \over {\partial s}}+{{iK}\over{R}}
\right) f_1 \eqno {39} $$
where $L'$ is as defined in eq. 23.
These equations are simply integrated over with the result
$$f={{-ivs^2}\over{4R}} (k_1^2-k_2^2)+i(k_1^2-k_2^2) \left({{is^2}\over
{4R^2}} +{{Ks^3}\over{24R^3}} \right). \eqno {39} $$
To get the Green's function at this order we operate on this function by
$\bf{L}$  given in eq. 31.  The result is seen to yield
$$G_F^(1) =[(x-x')^2-(y-y')^2]\left(A_1{{s^2v\Theta(v)-{s'}^2v'\Theta(v')}
\over{(s-s')[\quad   ]}}
+A_2{{\Theta(v)s^2+\Theta(v'){s'}^2}\over{(s-s')^2[\quad   ]}} \right. $$
$$\left.+A_3{{s^3\Theta(v)-{s'}^3 \Theta(v')}
\over {(s-s')^3[\quad   ] }} \right) \eqno {40} $$
where
$$ [\quad   ] = (u-u')(v-v')-{{uu'}\over {2}}
\left( (x-x')^2+(y-y')^2 \right), \eqno {41}$$
and $A_1, A_2, A_3$ are constants.
We see that this result is of the Hadamard form.  We find that all these
terms have the same type of ultraviolet singularity as the flat part.
We could not find a finite part of this expression.

If we go one order higher, we end up with the differential equation
$$L_2 g=v^2 \left( {{iRs}\over{2}}+3s^2(k_1^2+k_2^2)+(k_1^2-k_2^2)^2
\left({{-is^3}\over {4R}} \right) \right) $$
$$+v\left( {{-s}\over{2}}+{{iKs^2}\over {4R}}-
{{(k_1^2-k_2^2)^2s^3}\over{4R^2}}
+{{iK (K_1^2-k_2^2)^2s^4}\over{24R^3}} \right) \eqno {42} $$
when we make the ansatz $\phi_2=\phi_0 g$.  We take
$$g=v^2g_1(x,y,s)+vg_2(x,y,s)
+g_3(x,y,s). \eqno {43}$$
Going through similar steps we find
$$g_1= {{-s^2}\over{8}} +{{is^3(k_1^2+k_2^2)}
\over{2R}} +{{(k_1^2-k_2^2)^2 s^4}\over {32 R^2}}, \eqno {44} $$
$$g_2= {{-i3s^2}\over {8R}}-{{13Ks^3 }\over {12R^2}}+{{is^4}\over{32 R^3}}
\left( (k_1^2-k_2^2)^2 +K(k_1^2+k_2^2) \right) $$
$$ +{{s^5K}\over{160R^4}}(k_1^2-k_2^2)^2 ,\eqno {45} $$
$$g_3= {{3s^2}\over {8R^2}}-{{i55Ks^3}\over {16R^3}} -{{s^4}\over{32 R^4}}
\left( (k_1^2-k_2^2)^2+K(k_1^2+k_2^2)+{{13K^2}\over{3}} \right) $$
$$+{{is^5K}\over{320R^5}}\left(3(k_1^2-k_2^2)^2+K(k_1^2+k_2^2) \right)
-{{K^2s^6(k_1^2-k_2^2)^2}\over{1920R^6}} .\eqno {46} $$

We see that when the Green's function is calculated using these functions
we get no infrared divergences for $g_1$.
When we use $g_2$, we have to perform integrals of the type
$$ \int_0^{\infty} {{d \alpha} \over {\alpha}} exp \left[ {{-i(u-u')(v-v')}
\over {uu' \alpha}} \right] \eqno {47}$$
which results in a soft infrared divergence.  If we use an infrared
regulator mass $m^2$ ,  we get terms that go as
$log\left( (m^2(u-u')(v-v') \right) $.
$m^2$
cancels out when derivatives are taken to obtain $<T_{\mu \nu}>$.

The infrared divergences are more severe for $g_3$.
There we have to perform integrals of the type
$$ \int_0^{\infty} d\alpha exp \left[ -i{{(u-u')(v-v')}
\over {uu' \alpha}} \right] \eqno {48} $$
which are linearly divergent at the upper limit for the massless case.
If we use a massive field as an infrared cut-off, we get terms that go as
${{1}\over {m^2}}$ as $m$ tends to zero.  This term multiplies the
whole expression and does not drop out on differentiation.
\bigskip
\noindent{\bf{
IV.Going to de Sitter space }}

\noindent
Here we treat the same problem for an impulsive wave in de Sitter space.
It is known that an exact solution can be found for an impulsive
wave propagating in de Sitter space if we multiply the Minkowski space
solution by the conformal factor $\left( 1+{{\Lambda uv}\over {6}} \right)$
$^{/13  }$.  This is reflected to the $G_F$ by  factors multiplying the
expression from both sides.
$$ G_F^{S}= \left( 1+{{\Lambda uv}\over {6}} \right) G_F^{M}(x,x')
\left( 1+ {{\Lambda u'v'}\over {6}} \right),  \eqno {49} $$
where $G_F^S$ and $G_F^M$ are the de Sitter and Minkowski space Green
functions respectively.

Note that
$$\left(1+{{\Lambda uv}\over {6}} \right) \left(1+{{\Lambda u'v'}\over {6}}
\right)=\left(1+{{\Lambda  UV}\over {6}} \right)^2
+{{\Lambda}\over {12}} (u-u')(v-v') $$
$$ \left( {{\Lambda }\over {12}} \right)^2 \left( -(u-u')^2V^2-(v-v')^2U^2
+{{(u-u')^2(v-v')^2}\over {4}} \right), \eqno {50} $$
where $U={{u+u'}\over {2}}, V= {{v+v'}\over {2}} $.
Using this expression we see that for $G_F^{S(1)}$      we get terms
in the coincidence limit that give a well-defined finite expression for
$<T_{\mu \nu}^{(1)}>$.   Here the regularization is performed using the
point-splitting method, exactly in the same way one would do for the zeroth
part of the de Sitter universe.  One sees that for the warp function we
studied in Section II, we get a finite contribution.
If we use $h=(\zeta)^{1+\delta+i\epsilon}$ we get
$$<T^{(1)}_{vv} > = {{-\Lambda^2 u \delta(v)}\over {2(96)^2 \pi}} \left(
\delta log(x^2+y^2)-4\epsilon tan^{-1} {{y}\over {x}} \right). \eqno {51} $$
This result is finite, contrary to the case for propagating in Minkowski space.
This is no contradiction to our previous results, since this expression
vanishes as $\Lambda$ goes to zero.
One can also show that  $^{/14 } $ for $ h=
\left( {{\zeta-1}\over {\zeta+1}} \right)^{1+i \sqrt{2} \epsilon}$
we get
$$<T_{vv}^{(1)}>= {{-u\epsilon \Lambda^2}\over {(96)^2 \pi}}
\left( 2xtan^{-1} {{y}\over {x+1}} -2x tan^{-1} {{y}\over{x-1}}
-2y log \left( {{(x-1)^2+y^2}\over {(x+1)^2+y^2}} \right) \right) \delta(v).
\eqno {52} $$

All of these calculations are in first order perturbation theory.
We saw in Section III that there are problems in the second order
calculation due to the appearance of infrared divergences for the special
warp function used.  One can show that these infrared divergences are
generic for the Nutku-Penrose impulsive metric $^{/15}$ in second order
and all can be tamed to give a finite expression for $<T_{\mu \nu}>$
in the following way.

The  solution for the same type of wave 
in de Sitter space has the multiplicative factor $ {{\Lambda} \over {m^2}} $
multiplying a finite expression.  This fact suggests taking  the curvature
of the de Sitter space $\Lambda$ proportional to the infrared parameter
$m^2$.  These two terms both have the same dimensions, as were introduced
only as a technical aid to get rid of the ultraviolet and infrared
divergences.
Our end result should be independent of them.  This is achieved if we
take them to be proportional to one another, and let them go to  zero
at the same rate.  The finite proportionality constant that may appear
is absorbed in the perturbation expansion parameter, $\alpha^2$ for
the case studied in Section III.

Studying this particular case, one sees that applying the operator
${\bf{L}}$ as given in eq. 31 on   the expression $g_3$, eq.(46),
results in divergent terms of the form
$$ {{\Theta (v) s^2}\over {(s-s')^2 m^2}}
+{{\Theta(v') {s'}^2}\over {(s-s')^2 m^2 }} \eqno {53} $$
in Minkowski space.  If we go to de Sitter space, we have to multiply
this solution by the factor ${{\Lambda}\over{6}}(u-u')(v-v')$.  This operation
gives rise to a finite expression when both $\Lambda$ and $m^2$
are taken to zero at the same rate.

The result of this exercise is that by performing the calculation in
de Sitter space and taking the limit where the curvature of this space
goes to zero at the  same rate as the infrared parameter, we indeed get
$$ <T_{vv} > \propto  {{\delta(v)}\over {u^3}} \eqno {54} $$
which is proportional
to the sole non-zero component of the Riemann tensor , as anticipated
in the first place.

In the next section we see that this property is very special to the
impulsive spherical waves and is not shared with the spherical shock
or impulsive plane and impulsive pp waves.
\bigskip
\noindent {\bf{
V.Spherical shock wave}}

\noindent
In first order perturbation theory both the impulsive
and the shock wave cases showed similar behaviour.  The two solutions are
essentially different, though.  The shock wave solution has a dimensional
constant which is lacking in the impulsive wave solution.  Since
in quantum field theory , models
with dimensional and dimensionless constants belong to different classes,
we thought similar distinction
between these two models may exist.
Relying on these motivations we planned
to check whether there is a qualitative difference
between the two solutions exhibited by their behaviour at higher orders .

We will show that the infrared divergences which may be cancelled via a
detour in de Sitter space are  absent in the shock wave calculation.
Another point of difference is the importance attributed to the homogenous
solutions
in these two cases.
The homogenous solutions give just the free Greens function for the
impulsive case, whereas they result in a totally different contribution to the
Greens function in the shock wave.
This is an artifact of the presence of a dimensional coupling
constant in the latter case.
What may be more interesting is the fact that
just the contribution of the first order calculation
contributes to $<T_{\mu \nu}>$ in de Sitter space.
The higher order terms cancel out when the VEV of the stress-energy tensor is
computed.

We start with the metric  $^{/ 12 } $,
$$ ds^2= 2P du dv + 2u P_{\zeta} d\zeta dv +
2u P_{\overline{\zeta}} d{\overline{\zeta}}dv
-2u^2 d\zeta d\overline{\zeta}. \eqno {55} $$
Here we use the same variables as those given for the impulsive wave; i.e.,
$u$ is a Bondi-type luminosity distance,$v$ is a timelike coordinate
that can be regarded as retarded time and $\zeta$ and $\overline{\zeta}$ 
replace the angles on the stereographic projection of the sphere.  $P$ is
defined by 
$P={1\over {|h_{\zeta}|}}$,
where $h$ is an arbitrary function of the argument $\zeta+gv \Theta(v)$.
$g$ is the dimensional coupling constant, with dimensions of $mass$ 
and $\Theta$ is the Heavyside unit step function.  Nutku $^{/12}$
shows that the
identifying factor in this metric is one component of the Weyl tensor,
$$ \Psi_4= -{{1}\over {uP}} \left[ P P_{\zeta \zeta \zeta} 
- P_{\zeta \zeta} P_{\zeta} \right] \Theta(v). \eqno {56 } $$

When we write the  d'Alembertian operator in the this background metric,
we get 
$$ \sqrt{-g} \kutu=L= 2u^2 {{\partial^2}\over{\partial u \partial v}}+
2u {{\partial} \over
{\partial v}} + 2u P_{\zeta \overline{\zeta}} {{\partial} \over {\partial u}}
-2 u^2 {{P_{\zeta \overline{\zeta}}}\over {P}} {{\partial ^2}
\over {\partial u^2}}
-4u{{P_{\zeta} P_{\overline{\zeta}}}\over {P}} {{\partial} \over
{\partial u}}   $$
$$ +2u \left( P_{\overline{\zeta}} {{\partial} \over {\partial \zeta}}
+P_{\zeta} {{\partial}\over {\partial \overline{\zeta}}} \right)
{{\partial} \over {\partial u}}-2P {{\partial ^2} \over 
{\partial \zeta \partial \overline {\zeta}}}  \eqno {57} $$

In our particular case we take $h=(\zeta+gv\Theta(v))^{1+i\delta}$, where
$\delta<<1$ and is the expansion parameter.

We expand the operator $L$  in powers of $\delta$.
$$L=L_0+\delta L_1 + {\delta}^2 L_2+...  \eqno {58}$$
If we use real variables, $\zeta ={{x+iy}\over {\sqrt 2}}$, we get
$$ L_0= 2u^2{{\partial ^2}\over { \partial u \partial v}}+
2u{{\partial}\over {\partial v}}-{{\partial^2} \over {\partial x^2}}
-{{\partial^2} \over {\partial y^2}},
\eqno {59} $$
$$L_1= -{{2u}\over {x^2+y^2}} \left( y {{\partial}\over {\partial x}}
-x{{\partial} \over {\partial y}} \right) {{\partial } \over {\partial u}}
-tan^{-1}{y \over {x}} \left( {{\partial^2} \over {\partial x^2}}
+{{\partial^2} \over {\partial y^2}} \right) ,\eqno {60} $$
$$L_2=-{1 \over {x^2+y^2}} \left( u{{\partial} \over {\partial u}}+
u^2 {{\partial^2}\over {\partial u^2}}+ 2 tan^{-1} {y \over {x} }
\left(y{{\partial} \over {\partial x}} -
x{{\partial} \over {\partial y}} \right)
u {{\partial} \over {\partial u }} \right) $$
$$+{1\over {2} } \left( 1-(tan^{-1} {y \over {x}})^2 \right)
\left( {{\partial ^2}\over {\partial x^2}}+{{\partial ^2}\over
{\partial y^2}}
\right) . \eqno {61} $$
We expand both the solutions and the eigenvalue in terms of
$\delta$,
$$\phi=\phi_0+\delta \phi_1+\delta^2 \phi_2+..., \quad \lambda=\lambda_0+
\delta \lambda_1
+\delta^2 \lambda_2+...  .  \eqno {62} $$ 
Up to second order we get the coupled set of equations
$$ L_0\phi_0 =\lambda_0 \phi_0 , \eqno {63} $$
$$L_1 \phi_0+L_0 \phi_1=\lambda_1 \phi_0+\lambda_0 \phi_1 \eqno {64} $$
$$L_2 \phi_0+L_1 \phi_1+L_0 \phi_2= \lambda_2 \phi_0+ \lambda_1 \phi_1
+\lambda_0 \phi_2 \eqno {65} $$

We make the ansatz $\phi_1=f \phi_0$,
$\phi_2=h\phi_0.$  The ansatz for $\phi_1$ results in the equation
$$\left[ -2{{\partial^2}\over{\partial s \partial v}}
-2iR{{\partial}\over{\partial s}} +\left({{iK}\over {R}}{{\partial}\over
{\partial v}}\right)  -2i \left( k_1 {{\partial}\over{\partial z}}
+k_2 {{\partial} \over {\partial y}} \right)-{{\partial^2}
 \over {\partial z^2}} -{{\partial^2}\over {\partial y^2}} \right] f= $$
 $$ {{2i}\over {z^2+y^2}}(k_1y-k_2z) \left( {{iKs}\over {2R}}-1 \right)
 -(k_1^2+k_2^2) tan^{-1} \left({{y}\over {z}} \right) \eqno {66} $$
We recognize that when we plug these ansatze into our equations, we
get inhomogenous equations where on the right hand side the variable
$v$ always appears in the combination $z=x+gv$ for $v>0$.  $v$ is not
an independent variable in $f$ and $h$.  Thus we substitute to new
set of variables $u,z,y$.  This necessitates replacing all $v$ derivatives
by $g{{\partial}\over {\partial z}}$.  This makes an expansion in powers
of $g$  possible.
We take $$f=f_0(z,y,u)+gf_1(z,y,u),$$
$$ h=h_0(z,y,u)+gh_1(z,y,u). \eqno {67}$$
$\phi_0$ is given by, as in the previous cases
$$\phi_0= {{\exp[i(k_1x+k_2y+Rv-{K\over {2Ru}})]}\over
{u\sqrt{|R|} (2\pi)^2}}. \eqno {68} $$
where  $K,k_1,k_2 ,R$ are the
seperation constants which act as eigenfrequencies to be integrated
 over to find the Greens Function.
For $z=x+g \Theta(v) v, s={{1}\over{u}}$ we find
$${\cal L}_0 f_0 =I \eqno {69} $$
$${\cal L}_1 f_0+{\cal L}_1 f_0 =0 \eqno {70} $$
where
$${\cal L}_0= -2iR{{\partial}\over {\partial s}}-2i \left( k_1{{\partial}
\over{\partial z}}
+k_2{{\partial}\over{\partial y}} \right)-{{\partial^2}\over {\partial y^2}}
-{{\partial^2}\over {\partial z^2}} \eqno {71} $$
$${\cal L}_1= -2{{\partial^2}\over {\partial s \partial z}}+{{iK}\over {R}}
{{\partial}\over {\partial z}} \eqno {72} $$
$$ I={{2i}\over {(z^2+y^2)}} (k_1y-k_2z) \left[ {{iKs}\over {2R}}-1 \right]
-(k_1^2+k_2^2) tan^{-1} {{y}\over {z}} \eqno {73} $$
We see that if we use
the variables $z+iy$ and $z-iy$  the right hand side can be written as
a sum of functions of this two variables.  We  also write the differential
operator on the left hand side in terms of these new set of variables
and after some algebra we find
$$f_0=\left[{{Ks}\over{4R}}+{{i}\over{2}} \right] ln{{{z+iy}\over{z-iy}}}
-\left[{{K+k_1^2+k_2^2}\over{4}} \right]\left( {{1}\over{(k_1+ik_2)}}
\left[(z+iy) ln(z+iy) -(z+iy) \right]\right. $$
$$\left.-{{1}\over{(k_1-ik_2)}} \left[  (z-iy) ln (z-iy)-(z-iy) \right]
\right)  \eqno {64} $$
$$ f_1=\left({{iK}\over{2R}} +{{K^2 s}\over{8R^2}} \right)
\left({{ln(z+iy)}\over {k_1+ik_2}}-{{ln (z-iy)}\over {k_1-ik_2}} \right) $$
$$-{{K}\over {4R}} \left({{1}\over{2(k_1^2+k_2^2)}}+{{K}\over
{(k_1^2+k_2^2)^2}} \right)
\left((k_1-ik_2)^2 \left( (z+iy) ln(z+iy)-(z+iy) \right)\right. $$
$$ \left.-(k_1+ik_2)^2 \left( (z-iy) ln(z-iy) -(z-iy) \right) \right)
.\eqno{75} $$

The Green function in first order for the warp function chosen above
is given in reference 16.  The expression given in this paper was
calculated by a different method and also includes the contribution coming
from the homogenous solution.  Here we calculate them seperately .
We can show that we do not get a finite contribution to the vacuum
expectation value of the   stress-energy tensor. When the contribution only
from the inhomogenous equations (eq.s 69,70) is taken, the Green function  is
given by a product of  two factors one is which always equals zero when the
limit in $x=x', y=y'$
is taken without equating $u$, $v$ to $u',v'$ respectively.

In second order in $\delta$, we can reduce the differential equation to
the system
$${\cal L}_0 h_0 = M_0 ,\eqno {76}  $$
$${\cal L}_0 h_1 +{\cal L}_1 h_0 = M_1 , \eqno {77} $$
where
$${\cal L}_0 =-2iR {{\partial} \over {\partial s}} -2i \left( k_1
{{\partial}\over {\partial z}}
+k_2 {{\partial} \over {\partial y}} \right)
-{{\partial ^2} \over{\partial z^2}}-
{{\partial^2} \over {\partial y^2}}, \eqno {78} $$
$${\cal L}_1 =-2{{\partial ^2}\over {\partial s \partial z}} +{iK \over {R} }
{{\partial}\over {\partial z}}.\eqno {79} $$
Here $s={1\over{u}}$.  $M_0$ and $M_1$ are given as
$$M_0={1\over{2}}K-{1 \over {z^2+y^2}}
\left( 1-{3iKs \over {2R}}-{K^2s^2 \over {4R^2}} \right)
\left(1-2i (k_1y-k_2z) tan^{-1} {y \over {z}} \right)$$
$$+K\left( {5 \over {2}}-{iKs \over {2R}} \right)(tan^{-1} {y\over {z}})^2$$
$$-2K \left( \left({log (z^2+y^2) \over {2}}-1 \right)(k_1y-k_2z)
+tan^{-1}{y\over {z}}(k_1z+k_2y) \right) $$
$$ \times \left(\left(1-{iKs \over {R}}\right)
  {{k_1y-k_2z} \over {(z^2+y^2) K}}
-{i\over {2}} tan^{-1} {y\over {z}} \right)             $$
$$-{i \over {z^2+y^2} } \left( (1-{iKs \over {2R}})
\left( (k_1z+k_2y) log (z^2+y^2) +2(k_2z-k_1y) tan^{-1}
{y \over {z}} \right) \right)
\eqno {80} $$
$$M_1={{2iyK}
\over{(z^2+y^2)R}}\left(1-{{iKs}\over {4R}} \right) tan^{-1} {y \over z}
-{2k_1 \over {(z^2+y^2)R}}\left(1-{iKs \over {R}}-
{K^2s^2 \over {8R^2}}\right)$$
$$- \left( 2k_1 tan^{-1} {y\over z} -k_2 log(z^2+y^2) \right)$$
$$\left( \left({{-iK^2 s}\over {8R^2}}+{{5K}\over {4R}} \right)
tan ^{-1} {y \over z} - {{i(k_1y-k_2z)} \over {(z^2+y^2)R}}
\left( 1-{{iKs}\over {R}}-{{K^2 s^2} \over {8R^2}} \right) \right)$$
$$+\left((k_1^2-k_2^2) \left( 2z tan^{-1}{y\over z}+
y log(z^2+y^2)-2y \right)\right.$$
$$\left.
-2k_1k_2 \left( z log(z^2+y^2)
-2y tan^{-1} {y \over z} -2z \right) \right) $$
$$ \times \left( {3Ki \over {8R}} tan ^{-1} {y \over z}
-{3 \over {4(z^2+y^2)R}}(k_1y-k_2z)(1-{iKs \over {2R}}) \right)
-{3i \over {4(z^2+y^2)R}}
\left(1-{iKs \over {2R}} \right)$$
$$\times \left( (k_1^2-k_2^2)
\left(z log(z^2+y^2) -2y tan ^{-1} {y \over z} \right)
+2k_1k_2 \left( y log(z^2+y^2)+2z tan ^{-1} {y \over z} \right) \right)
\eqno {81} $$
In these expressions we took the `mass-shell' condition,
which is imposed in the
calculation of the Greens function; i.e. we set $k_1^2+k_2^2$ equal to $K$.
One can
check that after we perform the $ K$ and $k_1, k_2$ integrations
the effect of these two expressions are exactly the same.

We see that, contrary to the impulsive wave calculation,
in both $M_0$ and $M_1$, there are no terms that are independent
of $z$ and $y$ except a single term which is proportional to $K $.
To be able to obtain terms in $<T_{\mu \nu}>$
that diverge as the infrared parameter goes to zero, we need inverse powers
of $R$ which are not multiplied by $K$ or $k_1^2, k_2^2$.
Each inverse power of $R$ means a higher infrared divergence,
resulting in terms that go  $m^2,1, \log m^2,{1\over {m^2}},
{1\over {m^4}}$, etc...as the power of ${{1}\over{R}}$ is increased,
whereas each power of $ K,k_1^2,k_2^2$ means one lower order in the same
divergence. In free space the power of $m$ is zero.
There is no  infrared divergence.

In reference 15, we generated these divergences at second order and then
cancelled them with the cosmological constant of the de Sitter solution.
Our mechanism for obtaining these infrared divergences was as follows.
We isolated the $ -2iR {{\partial} \over { \partial s}}$
in the operator ${\cal L}_0 $ from the others and equated it to the
term which did not contain $ z $ or $y$.
$$-2iR {{\partial} \over {\partial s}} h'_0 =cs  \eqno {82}$$
where $c$ can be a function of $v$ but not that of $z$ and $y$.  Then
$h'_0 = {{ics^2}\over { 4R}}$  has an extra power of $1/R$
compared to the other terms.
The second iteration gives us $h'_1 \propto s^3/{R^2} $.
Such a term will induce $1/m^2$ factor in the expression for
the Greens Function , $G_F$, and this infrared
mass will be retained in $<T_{\mu \nu}>$.

For the shock wave solution all the terms in $M_0$ and $M_1$ are either
functions of $z$ and $y$ or are multiplied by $K$.  We can not isolate a part
of the operator ${\cal L}_0 $ and equate it to a single term on the RHS.
We may still study the shape of solutions.  Let us make the ansatz 
$$ h_0= s^2 h_{00}+sh_{01}+h_{02} .\eqno {83} $$
We find that
$$h_{00}=-{{4K^2}\over {32R^2}} \left(tan^{-1} {{y}\over{z}}\right)^2;
\eqno {84} $$
$$h_{01}=
-{{3iK}\over {4R}} \left(tan^{-1} {{y}\over {z}}\right)^2 \eqno {85}$$
etc.  These solutions show that we are not getting powers of $R$ 
in the denominator with no powers of $K$  in the numerator.  
This fact makes it clear that there is no
way of acquiring   ${{1}\over {m^2}}$ in the Green function, since we need
an extra  $R^{-2}$ factor, which necessitates at least one power in the
first integration.
On RHS
only the combination ${K^2 \over {R^2}} $ and ${K \over {R^2}} $ exist .
${K^2 \over {R^2}}$ gives exactly
the singularity structure as the free case, and ${K \over {R^2}}$ gives a
logarithmic divergence
which is cancelled in the $<T_{\mu \nu}>$ calculation.

At this point we note that we can find solutions of equations 76 and 77
even if
$M_0$ and $M_1$ are set to zero.  These are the homogenous solutions of the
problem which give a non trivial contribution for the shock wave calculation.
These solutions of the homogenous equations generated the only non-zero
part of the Green function when we took $x=x', y=y'$ in the first-order
calculation.
Since $M_0$ and $ M_1$ are independent of $v$ we can assume a powers series
expansion in $v$ for a chosen order in $g$.  For the sake of illustration we
take a solution in third order in $g$ and write the expansion as
$$ f^{(1,3)}_H = g^3 ( v^3 f^{(1,3)}_{1H} + v^2 f^{(1,3)}_{2H}
+v f^{(1,3)}_{3H}+f^{(1,3)}_{4H}). \eqno {86} $$
Here $f_{1H}^{(1,3)}(s,z,y)$ has dimension zero, $f_{2H}^{(1,3)}(s,z,y)$
has dimension
minus one, etc.. Inverse powers of $v$ are excluded by the regularity
at $v=0$.  One can show that taking powers of
$v$ higher than that of $g$ do not give  results that differ from the
free case.  A similar expansion in the impulsive case would go as
$$ f_{H}^{(1,3)}=(v^3R^3 f_{1H}^{(1,3)}+v^2R^2 f_{2H}^{(1,3)} +...)\eqno{87}$$
where $f_{1H}^{(1,3)}$ etc., have the same dimensions as above,
since the only
free dimensional parameters are $v$ and $R$.  This gives the result of the
flat ,Minkowski, space Green function.

Keeping track of
powers of $v$ we get a system of four equations.  We note that the first of
these equations
$${\cal L}_0 f^{(1,3)}_{1H} = 0 \eqno {88} $$
has a solution for any function $$F=F({s \over {R}}(k_1 \pm i k_2)
-(z \pm iy)).\eqno {89}$$  We can also show that the singularity behaviour of the
Greens Function is independent of the form of $F$.  At the end we get,
 for the worse infrared poles the expressions
$$ G_H^{(1,H)} =2 \pi c_1 \left( {{v^3 \Theta(v) + v'^3 \Theta(v')}\over
 {(u-u')(v-v')}} \right) ,\eqno {90} $$
$$ G_H^{(2,H)} = 2 \pi c_2 \left( {{ uv^2 \Theta(v) + u'v'^2 \Theta(v')}
 \over {(u-u')^2}} \log (2m^2(u-u')(v-v')) \right), \eqno {91} $$
$$ G_H^{(3,H)} = 2 \pi c_3 \left( {{ u^2v \Theta(v)+u'^2v' \Theta(v')}
\over {(u-u')^4 m^2}} \right)   , \eqno {92} $$
$$ G_H^{(4,H)} = 2 \pi c_4 \left( {{ u^3 \Theta (v)+u'^3 \Theta (v')}
\over {(u-u')^6 m^4}} \right). \eqno {93} $$
Here $c_i$ are functions of $x$ and $ y$, depending on the form for $F$ used.
$m$ is the infrared mass.
If we use a linear functionfor $F$,  $c_i$ is proportional to $y$ or $x$.

Upon symmetric differentiation with respect to $v$ and $v'$, the terms with ${1 \over {m^2}}$ and
$ {1 \over {m^4}} $ vanish.  We  find a finite contribution only if we
go to the  de Sitter space, i.e. multiply by the factor
$(1+{\Lambda uv \over {6}})(1+{\Lambda u'v' \over {6}}) $.
In this case we get
$$ <T_{vv}> = g^3 \left(  f_1 \Lambda v \Theta(v) +
f_2 \Lambda^2 uv^2 \Theta(v) \right) \eqno {94} $$
which goes to zero with $\Lambda$ when we go back to Minkowski background.
In this expression $f_1$ are regular functions of $x,y$.
One can also show that any terms with less diverging powers of $m^{-2}$
in $G_H^{(4,H)}$ do not give a finite contribution even in
de Sitter background.

Here we tried to show that two qualitative differences exist
between the shock and the
impulsive wave solutions proposed by the same group $^{/10,12}$.
In the shock wave
solution  the infrared divergences which may be used to tame the
ultraviolet divergences
to  result in finite contributions to $<T_{vv}>$ are absent in second order
perturbation theory.  We can not find finite contributions
to $<T_{v v}>$ in Minkowski
space.  If we go to de Sitter space, though, we get a finite contribution
which is proportional to $\Theta$ function,
which is the signature of a shock wave solution.

The homogenous solutions, in the shock wave,
give contributions to the Greens function expression which are different
from the free case.  These solutions also give a finite contribution to
$<T_{vv}>$ in de Sitter space.  The presence of these nontrivial solutions
is only due to the dimensional coupling constant.  The presence of $g$
in the expansion makes it necessary to have an extra power of ${1\over {R}}$
in the solution which results in a nontrivial term in $G_F$.
\bigskip
\noindent
{\bf{
VI.  Plane impulsive wave}}

\noindent
As the last example here we take the metric describing an impulsive plane
wave $^{/17}$    ,
$$ds^2=2dudv-|d {\overline {\zeta}} + q_{\zeta \zeta} v \Theta(v) d\zeta|^2 .
\eqno {95} $$
If we take $ q=g{{\zeta^2}\over {2}}$ we get a plane wave.  If the power is
higher than quadratic we get pp waves $^{/18 }$.
The d`Alembertian  operator in this metric is written as
$$ L= 2 \partial_{u} \partial_{v} -{{2vg^2}\over{1-v^2g^2}} \partial_{u}
-{{1}\over{(1+vg)^2}} \partial_{x}^2 -{{1}\over {(1-vg)^2}}
\partial_{y}^2 \eqno {96} $$
If we expand up to second order in the coupling constant $g$, we get
$$L \approx 2 \partial_{u} \partial_{v} -2vg^2 \partial_{u}
-(1+3(vg)^2)(\partial^2_{x}+\partial^2_{y}
+2vg (\partial^2_{x}-\partial^2_{y} )). \eqno {97} $$
The zeroth-order solution is similar to the previous cases, eq. (15) resulting
in a Green function that goes as
$$ {{A}\over {(u-u')(v-v')-{{1}\over {2} }[(x-x')^2+(y-y')^2]}} \eqno {98} $$
for constant $A$.  We expand the solution in powers of $g$ and take the first
order solution
as $ \phi^{(1)}= f \phi^{0}$. It is straightforward to solve for $f$ and 
we get
$$ f={{(k_1^2-k_2^2)u}\over {2iR}} \left[v+{{i}\over {R}} -{{Ku}\over {4R^2}} \right]
\eqno {99} $$
For the second order solution we again take $\phi^{(2)}= \phi^{(0)} h $.
Here $ h= v^2h_1(x,y,u)+vh_2(x,y,u)+h_3(x,y,u) .$
A straightforward calculation gives us
$$ h_1={{3i}\over {2R}} (k_1^2+k_2^2) u
-{{u^2}\over {4R^2}}(k_2^2-k_1^2)^2 ,\eqno {100} $$
$$ h_2={{u}\over{R^2}}({{K}\over {2}}-3(k_1^2+k_2^2))-{{3iu^2}\over{4R^3}}
\left( (k_1^2-k_2^2)^2+K(k_1^2+k_2^2) \right)+
{{K(k_1^2-k_2^2 )^2 u^3}\over {8R^4}}, \eqno {101} $$
$$ h_3= -i{{u}\over {R^3}}({{K}\over{2}}-3(k_1^2+k_2^2))+{{u^2}\over {R^4}}
({{1}\over{8}} \left(3(k_1^2-k_2^2)^2+3K(k_1^2+k_2^2)-K^2 \right)$$
$$+{{iu^3}\over{8R^5}} \left(2K(k_1^2-k_2^2)^2+K^2(k_1^2+k_2^2) \right)
-{{K^2(k_1^2-k_2^2)^2 u^4}\over{64R^6}}. \eqno {102} $$
Here we see that a peculiar thing happens.  We almost get a finite expression,
but we have one power of $(u-u')$ too many in the denominator.
If we go to de Sitter space to cancel both the ultra-violet and
infrared divergences, we see that the finite part of $<T_{vv}>$ goes as
$$ <T_{vv}>\propto -2 {{\Lambda^2}\over {m^2}} \Theta(v)  \eqno {103} $$
which is finite only in  de Sitter space.  One power of the curvature
cancels with the infrared parameter since we take $ \Lambda \propto m^2 $
, but the remaining power takes the contribution to zero when we go back to
Minkowski space.

This result which is in accord with general arguments of Deser $^{/6,7 } $
is a check that our method does not contradictany known results.

One can show that this result does not change in the presence of a pp-wave
background.  Whether a wave is plane or pp type  depends only on the
form of the function $q(\zeta)$ in the metric.
The general behaviour of the expression for the vacuum expectation
value of the stress-energy tensor does not depend on the form of the
function $q$.  This form only changes an overall factor which can not decide
whether the whole expression is finite or null.  The same behaviour was
already seen in the different warp functions we have used for the spherical
wave.
\bigskip
\noindent
{\bf{CONCLUSION}}

\noindent
Here we reviewed the work on the vacuum fluctuations for the stress-energy
tensor for  a scalar particle in the background of different impulsive and
shock wave metrics.  We found a finite fluctuation only for the spherical
impulsive wave metric, whereas our result was null for the spherical
shock and plane impulsive waves.
We attribute the difference in the two cases to the presence of the
dimensionless and dimensional constants.

If we calculate the fluctuations
for a conformal metric, fluctuations should be absent $^{/2}$.
We  first perform perturbation theory about the Minkowski space, and
our perturbations are not strong enough to overcome the restrictions
imposed by conformal symmetry.  If we go to de Sitter space, and
perform perturbation around that metric, we do not have this obstruction.  
We always find finite fluctuations in that metric.  Note that de Sitter
space also tames our ultra-violet divergences.
It turns out that if we have dimensional coupling constants, we have
more severe ultra-violet divergences which are tamed only with having
higher powers of the curvature scalar of de Sitter space, multiplying
our expressions for the fluctuations.  This happens in the two latter
metrics, plane and pp waves, we have studied  .  The resulting infrared
divergences are not strong enough to cancel these scalar curvature
factors to give us a finite result.

In the spherical shock wave solutions, we could not generate such
divergences in the inhomogenous case.  For the homogenous solutions,
the ultra-violet divergences
could not be tamed by going to de Sitter space 
.  Cancellation of the infrared divergence , say in eq.92, by cancelling
it by the $\Lambda$ term was not sufficient to obtain a finite expression.

The presence of non-trivial homogenous solutions was another
novel feature of the shock waves.  It was amusing that only one of them
survived in the expression for $<T_{\mu \nu} > $.  If many of them
survived, we would have an ambiguity in defining this quantity.
It was also interesting that no matter how many orders in the coupling
constant one expands, one gets one and only one surviving term, the one
obtained already in first order.
\bigskip
\noindent
{\bf{
Ackowledgement:}}
\noindent
We thank Prof.Dr.Y. Nutku for his support and giving his
metrics prior to publication and for many discussions.  We thank Prof. A.N.
Aliev for illuminating discussions.  This work is
partially supported by T\" UBITAK, the Scientific and Technical Research
Council of Turkey and  M.H.'s work also by T\" UBA,
the Turkish Academy of Sciences.

\bigskip
\noindent
REFERENCES
\item {1.}  One may look any of the reviews present.  A manuscript
that appeared a few days ago depends on the talk given by Gibbons
at the  GR15 meeting at Poona, India:
G.W.Gibbons, " Quantum Gravity/String/M-theory as we approach the
3rd Millennium", Cambridge Univ. preprint, gr-qc/9803065 18 Mar 1998;

\item {2.} N.D.Birrell and P.C.W. Davies, {\it{ Quantum Fields in
Curved Space}}, Cambridge University Press, Cambridge 1982;

\item {3.}  S.A.Fulling, {\it{ Aspects of Quantum Field Theory in
Curved Space-Time}}, Cambridge University Press, Cambridge 1989;

\item {4.}  R.M.Wald, {\it{ Quantum Field Theory in Curved Spacetime
and Black Hole Thermodynamics}}, The University of Chicago Press,
Chicago , 1994;

\item {5.} C.I.Kuo and L.H.Ford, Phys. Rev. {\bf{D47}} (1993) 4510; 

\item {6.} T.M.Helliwell and K.A.Konkowski, Phys. Rev. {\bf{D34}} (1986) 1918,
B.Linet, Phys. Rev. {\bf{D33}} 1836, Phys.Rev {\bf{D35}} (1987)   536,
A.G.Smith in {\it{ The Formation and Evolution of Cosmic Strings}}, ed.
by G.W.
Gibbons, S.W.Hawking and T.Vachaspati, Cambridge University Press, Cambridge,
1990,p. 263 (formerly Tufts University preprint 1986;

\item {7.} S. Deser, J. Phys. A. {\bf{8}} (1975)  1972;

\item {8.} G.W.Gibbons, Commun. Math. Phys. {\bf{45}} (1975) 191;

\item {9.}   M.Horta\c csu, Class. Quant. Grav. {\bf{7}} (1990) L165, Erratum
: Class. Quant. Grav. {\bf{9}} (1992) 799,
M.Horta\c csu, J. Math. Phys., {\bf{34}} (1993) 690,
M.Horta\c csu and R. Kaya, {\bf{35}} (1994) 3043, Errata in J. Math. Phys.
{\bf{37}} (1996) 4199, M.Horta\c csu, Class. Quant. Grav. {\bf{13}} (1996) 2683,
Y.Enginer, M.Horta\c csu and N. \" Ozdemir, Int. J. Mod. Phys. A,
{\bf{13}}(1998)1201;

\item {10.} Y. Nutku and R. Penrose, Twistor Newsletter {\bf{34}} (1992) 9;

\item {11.} M.Horta\c csu,Class. Quant. Grav. {\bf{9}} (1992) 1619,
N.\" Ozdemir and M.Horta\c csu, Class. Quant. Grav. {\bf{12}} (1995) 1221,
M.Horta\c csu and K.\" Ulker, Class. Quant. Grav. {\bf{15}} (1998)1415;
-
\item {12.} Y.Nutku, Phys. Rev D {\bf{44}} (1991) 3164;

\item {13.} P.A.Hogan, Phys. Lett.A {\bf{171}} (1992) 21;

\item {14.} Y. Enginer et al, cited in reference 9;

\item {15.} M.Horta\c csu, 1996, cited in  reference 9;

\item {16.} N.\" Ozdemir et al, cited in reference 11;\

\item {17.} Unpublished work of Y. Nutku; I am grateful to him giving me
this information last year.

\item {18.} Unpublished work of Y. Nutku; I am grateful to him giving me
this information last year.

\end